\documentclass[reqno]{article}
\usepackage{ae} 
\usepackage[T1]{fontenc}
\usepackage[ansinew]{inputenc}
\usepackage{amsmath}
\usepackage{amssymb}
\usepackage{graphicx}
\usepackage{color}
\usepackage{epsfig}

\newcommand{\beq}{\begin{equation}}
\newcommand{\eeq}{\end{equation}}
\newcommand\beqa{\begin{eqnarray}}
\newcommand\eeqa{\end{eqnarray}}
\newcommand\bea{\begin{array}}
\newcommand\eea{\end{array}}
\newcommand\ba{\begin{array}}
\newcommand\ea{\end{array}}
\newcommand{\nn}{\nonumber}

\newcommand{\neqa}{\nonumber\end{eqnarray}}
\newcommand{\la}[1]{\label{#1}}
\newcommand{\vrho}{\varrho}

\newcommand{\eq}[1]{eq.(\ref{#1})}
\newcommand{\eqs}[2]{eqs.(\ref{#1},\ref{#2})}
\newcommand{\Eq}[1]{Eq.(\ref{#1})}

\newcommand{\Tr}{{\rm Tr}}

\newcommand{\half}{\frac{1}{2}}

\renewcommand{\d}{\partial}
\renewcommand{\O}{{\cal O}}

\newcommand{\re}{\relax{\rm I\kern-.18em R}}

\renewcommand{\sp}{p\hspace{-.40em}/}

\def\su2{{SU(2)}}

\def\a{{\alpha}}

\def\({\left(}
\def\){\right)}
\def\[{\left[}
\def\]{\right]}

\begin{document}

\renewcommand{\thefootnote}{\fnsymbol{footnote}}
\setcounter{footnote}{0}

\thispagestyle{empty}
\begin{flushright}
LPTENS-05/33\\
\end{flushright}
\vspace{.5cm}
\setcounter{footnote}{0}
\begin{center}
{\Large{\bf Double Scaling  and Finite Size Corrections   in  $\mathfrak{sl}(2)$ Spin Chain\par}
   }\vspace{4mm}
{\large\rm Nikolay~Gromov$^{a,b}$, Vladimir~Kazakov$^{a,\!}$\footnote{Membre de
l'Institut Universitaire de France}\\[7mm]
\large\it\small ${}^a$ Laboratoire de Physique Th\'eorique\\
de l'Ecole Normale Sup\'erieure et l'Universit\'e Paris-VI,\\
Paris, 75231, France\footnote{
\tt\noindent gromov@thd.pnpi.spb.ru, gromov@lpt.ens.fr\\
 \indent\ \ kazakov@physique.ens.fr}\vspace{3mm}\\
\large\it\small ${}^b$ St.Petersburg INP,\\
 Gatchina, 188 300, St.Petersburg, Russia
}

\end{center}
\noindent\\[20mm]
\begin{center}
{\sc Abstract}\\[2mm]
\end{center}

  We find explicit expressions for two first finite size
  corrections to the distribution of Bethe roots, the
  asymptotics of energy and high conserved charges in the $\mathfrak{sl}(2)$
  quantum Heisenberg  spin chain of length $J$ in the thermodynamical
  limit $J\to\infty$ for low lying states with  energies
  $E\sim 1/J$. This limit was recently studied in the context of
  integrability in perturbative ${\cal N}=4$ super-Yang-Mills theory.
  We applied the double scaling
  technique to Baxter equation,  similarly to the one used for large
  random matrices near the edge of the eigenvalue distribution. The
  positions of Bethe roots are described near the edge by zeros of  Airy function.
 Our method can be generalized to any  order in $1/J$.
 It should also work  for other quantum integrable models.

\newpage

\setcounter{page}{1}
\renewcommand{\thefootnote}{\arabic{footnote}}
\setcounter{footnote}{0}

\section{Introduction}
We study in this paper the integrable periodic Heisenberg  $XXX_{s}$ chain
of noncompact quantum spins transforming under the representation
$s=-1/2$ of $\mathfrak{sl}(2)$,
 in the so called thermodynamical limit of large  $J$, where $J$ is the length of the chain, in the ferromagnetic regime of low energies $E\sim 1/J$.
\footnote{It is  different from a more traditional regime
$E\sim J$  widely studied since many years, especially in the condensed matter  literature.}

The problem is known to be  solvable by the Bethe
ansatz approach (see for example \cite{Faddeev}) and  the energy of a state of $S$ magnons in dimensionless units is given by a simple formula
\beq\label{ENE} E=\sum_{k=1}^S \frac{1}{u_k^2+1/4}, \eeq
where the Bethe roots $u_j,\ j=1,2,\cdots,S$, parametrizing the momenta of magnons,
are solutions of a system of polynomial Bethe ansatz equations (BAE)
\beq\label{BAE} -\left(\frac{u_j-i/2}{u_j+i/2}\right)^J
=\prod_{k=1}^S\frac{u_j-u_k+i}{u_j-u_k-i},\;\;\;\;\;j=1,\dots,S. \eeq
It can be proven  that for this model the roots are always real.

 In the thermodynamical limit we will also consider a large
number of magnons $S\sim J$. It is clear then from \eq{ENE} that in
order to focus on the low lying states with energies $E\sim
1/J$ we should take the characteristic Bethe roots of the order
$u_j\sim J$. It means that the chain is very long and the spins are
rarely changing along it. The typical length of  spin-waves
(magnons) is of the order of the length $J$. Our goal is to study
the limiting $J\to\infty$ distributions of Bethe roots and the finite volume $1/J$
corrections to these distributions, to the energy and higher conserved charges.
In the  main order this thermodynamical limit for the compact Heisenberg $XXX_{1/2}$
chain of $\mathfrak{su}(2)$ spins  was already considered in
\cite{Sutherland}, and  later in \cite{Beisert:2003xu} in relation to the integrable  dilatation hamiltonian  in planar perturbative superconformal  ${\cal N}=4$ super-Yang-Mills
(SYM) theory. Its description and the general solution
in terms of algebraic curves was proposed in \cite{KMMZ} for the
$\mathfrak{su}(2)$ case\footnote{Following a similar approach of
\cite{ReshetikhinSmirnov} to a somewhat different limit of large
spin} and in \cite{Korchemsky:1997ve,Kazakov:2004nh}  for the   $\mathfrak{sl}(2)$ chain.
We will concentrate in the current  paper on  this last  case.
Generalization to the $\mathfrak{su}(2)$ case is straightforward.

The study of $1/J$ corrections in these systems was started recently
in the papers
\cite{Beisert:2005mq,Hernandez:2005nf} for the
simplest single support, or one cut distribution, whereas a similar
quantum $\hbar$ correction to the classical KdV solitons was already
found earlier in the general multi-cut case in \cite{Smirnov}.

The main results of our paper are:

1. The explicit formulas for the $1/J$ and $1/J^2$ corrections to
the general multi-cut distribution of Bethe roots and to the
corresponding energy of a Bethe state in terms of the underlying
algebraic curve.

2. The universal description of the distribution of Bethe roots in the
vicinity of an edge of a support in terms of zeroes of the Airy
function, similar to the double scaling limit in the matrix models.

3. Asymptotics of conserved local charges $Q_n(S,J)$
in the large $n$
limit.

4. Asymptotics of conserved global (non-local)  charges of high order.

Unlike the papers \cite{Beisert:2005mq,Hernandez:2005nf} using the method of singular integral equation corrected by  so
called anomaly term\footnote{This phenomenon of anomaly, or the
contribution of close eigenvalues in the thermodynamical limit of
BAE was first observed in \cite{Beisert:2005di}}, we will use here
the exact Baxter equation written directly for the analytical function - the resolvent of the root distribution (similar approach was used in \cite{Korchemsky:1997ve}). This equation is valid before any limit.

These results might be interesting for different  kinds of specialists.

First, for those who are studying  large N  random matrix or random
partition ensembles. In particular, the distributions of eigenvalues
in matrix models are described by similar (although not the same)
algebraic curves. The $1/J$ corrections remind the $1/N$
corrections, and the Airy edge distribution observed in our work is
known to describe the edge behaviour in the  double scaled matrix
integrals as well \cite{BrezinKazakov}. No doubt that by choosing
particular solutions of \eq{BAE} or modifying its l.h.s. (by
considering inhomogeneous chains) we can also find here various
multi-critical phenomena similar to those found in the matrix models
in the context of quantum 2D gravity applications \cite{Kazakov:1989bc}.

Second, it might be interesting to those who work on
various aspects of  AdS/CFT correspondence  in supersymmetric string
 and gauge theories and in particular on the integrability in ${\cal N }=4$
 SYM theory (see \cite{Beisertthese} and references therein). They might shed some
light on   quantum corrections to the classical limit of the AdS
dual of SYM,  the superstring on the $AdS_5\times S_5$ background,
now known only for particular classical solitonic solutions
\cite{FrolovTseytlin} of rotating strings. AdS/CFT correspondence
should manifest itself as the coincidence  of such corrections in
these so different integrable systems.  Their similarity, and even the coincidence in a
certain regime, was already observed on particular string and chain solutions, having
only one support for the Bethe roots distribution
\cite{Beisert:2005mq,Hernandez:2005nf,Schafer-Nameki:2005tn,Beisert:2005bv,Beisert:2005cw,Schafer-Nameki:2005is,Minahan:2005mx,Minahan:2005qj}.
1/J corrections were first studied for BMN states in \cite{Minahan:2002ve}, where
the integrable spin chain for ${\cal N}=4$ SYM was first proposed,
and then in \cite{Arutyunov:2004vx}.
The Airy edge behavior also
seems to be universal enough to manifest itself on both sides of AdS/CFT duality.
This behaviour observed in the present paper for  spin chains describing the spectrum
of anomalous dimensions in the perturbative SYM in thermodynamical limit, is natural
to  expect also in the quantized
string theory where the  classical limit, the analogue of the thermodynamical limit,
is described by similar algebraic curves obtained by the finite gap method
\footnote{ See \cite{Beisert:2005di,Beisert:2004ag} for this approach on both sides of duality}.
The universal edge behaviour might be completely driven by the similarity of integrable
structures of these two seemingly different systems.

The paper is organized as follows. In section 2 we summarize the explicit formulas for the Hamiltonian  and the local
and non-local conserved charges of the model. In section 3 we review the general solution of the model in thermodynamical limit when $J\rightarrow \infty$ and $E\sim 1/J$.
In section 4 we calculate the first  $1/J$ correction to the general multi-cut solution  from Baxter equation.
In section 5 we apply
the method of the double scaling limit to the Baxter equation and find the near branch point behavior.
In section 6 we combine the results of section 4 and 5 to find the explicit formula
for the  second, $1/J^2$ correction. In section 7 we give the formulas for $1/J$ corrections to the energy and compute the asymptotics of high conserved
local and non-local charges in all orders in $1/J$. Conclusions are devoted to unsolved problems and perspectives, as well as to the discussion of  parallels with the matrix models.
In appendix A we write explicit formulas for
$1/J$ expansion of BAE and in Appendix B we express  $1/J^2$ corrections
to the energy  for one-cut solution through certain double  sums in mode numbers, generalizing the results of \cite{Beisert:2005mq}.

\section{Hamiltonian, Transfer-matrix and Higher Charges of $\mathfrak{sl}(2)$ chain}

The
 hamiltonian of interaction of the neighboring spins $s_l,s_{l+1}$  can be
  written in an explicit  way \cite{Beisertthese}
\beq  H_{-1/2}=\sum_{l=1}^J  \hat H^{l,l+1}_{-1/2} \label{HH}  \eeq
with the Hamiltonian density
\beq \hat H^{l,l+1}_{-1/2}|k,m-k\rangle= \sum_{k'=0}^m \(
\delta_{k=k'}\(h(k)+h(m-k)\)-\frac{\delta_{k\ne
k'}}{|k-k'|}\)|k',m-k'\rangle, \eeq
where $\vert k_1,\ldots,k_l,k_{l+1},\ldots,k_J\rangle$ is a state
vector labeled by $J$ integers $k_j$ ($s=-1/2$ spin components) and
 $h(k)=\sum_{j=1}^k\frac{1}{j}$ are harmonic numbers.

The total momentum $P(u)$
\beq  e^{iP(u_j)}=\frac{u_j-i/2}{u_j+i/2}  \eeq
satisfies the (quasi-)periodicity condition following directly from \eq{BAE}
\beq\la{QPC}   P_{tot}=\sum_{j=1}^S\ P(u_j)=2\pi k/J,\quad k\in {\bf
Z.} \eeq
 In some applications, such as the spectrum  of anomalous
dimensions of operators
\footnote{ The operators of
the type $\Tr\(\nabla^{k_1} Z\cdots\nabla^{k_J}Z\)$ in SYM,
where $\nabla=\partial +A$ is a covariant derivative in a null direction and $Z$
is a complex scalar, represent the
state vectors $\vert k_1,\ldots,k_l,k_{l+1},\ldots,k_J\rangle$
 and the
dilatation hamiltonian is given at one loop  by the  $XXX_{-1/2}$ hamiltonian. }
in ${\cal N}=4$ SYM theory, we select  only purely
 periodic Bethe states
\beq  \la{periodic} P_{tot}=2\pi m,\quad m\in {\bf Z}. \eeq

We can also study other physically interesting quantities of this model, such as
 the local conserved charges $\hat Q_r$. They are defined as
  follows
\beq\la{CHAGEN} \hat T(v)= \exp \(i\sum_{r=1}^\infty \hat Q_r v^{r-1}\),  \eeq
where  the quantum transfer matrix
$\hat T(v)\equiv \hat T(v;0,0,\cdots,0)$
 is a particular case of the inhomogeneous transfer matrix
\beq\la{INHT}  \hat T(v;v_1,\cdots,v_J) = \Tr_{0}
\[\hat R_{0,1}(v-v_1)\cdots \hat R_{0,J}(v-v_J)\]   \eeq
and $\hat R_{0,j}$ is the universal $\mathfrak{sl}(2)$ R-matrix defined as \cite{Kulish:1981gi}
\beq \hat R_{0,1}(v)=\sum_{j=0}^\infty R_j(v) {\cal P}^{(j)}_{0,1},\qquad  R_j(v)=\prod_{k=1}^j\frac{v-ik}{v+ik}  \eeq
with ${\cal P}^{(j)}_{01}$
being the operator  projecting the  direct product of two neighboring spins  $s_0=s_1=-1/2$
to the representation $j$. Recall that
\beq\la{TTCOM} \[\hat T(v;v_1,\cdots,v_J),\hat T(v';v_1,\cdots,v_J)\]=0
\eeq
 for any pair
$v,v'$, due to Yang-Baxter equations on the $\hat R$-matrix.

The direct calculation shows that  $\hat P_{tot}=- \hat Q_{1}$
is the operator of the momentum, such that  and
$\hat H_{-1/2}=\hat Q_2$ is the hamiltonian \eq{HH}, etc. Those charges are local, in the sence that the charge density  of $Q_k$ contains $\le k$ consecutive spins.

There is a more  efficient way to generate higher charges than using \eq{CHAGEN}. We can use the recurrence relation
(see \cite{Beisertthese}):
\beq  r \hat Q_{r+1}=[B,\hat Q_r],   \eeq
where $B=i\sum_{l=1}^J  l\ \hat H^{l,l+1}_{-1/2}$ is the boost
operator. The
last formula is true up to some boundary terms destroying the
periodicity, which should be dropped.

Due to the integrability manifestly expressed by \eq{TTCOM} all
these charges commute and their
eigenvalues on a  Bethe state characterized by a set of Bethe
roots satisfying \eq{BAE} (enforcing the periodicity
of the chain or the quasi-periodicity of the Bethe state) are
given by \cite{Faddeev:1994zg}
\beq \la{localDef}  Q_r=\sum_{j=1}^S\
\frac{i}{r-1}\(\frac{1}{(u_j+i/2)^{r-1}}-\frac{1}{(u_j-i/2)^{r-1}}\).
\eeq

We can also study the non-local charges. They can be defined in many different ways.
The  definition could be similar to \eq{CHAGEN}, but  the expansion of $T(u)$ goes around $u=\infty$.
However, the most natural charges are defined through the resolvent of Bethe roots:
\beq\la{REST}    G(x)= \sum_{k=1}^S \frac{1}{xJ-u_k} =\sum_{n=1}^\infty d_n x^{-n}. \eeq
The $j$-th charge is the $j$-th symmetric polynomial of Bethe roots
\beq\la{OURCH}   d_n=\frac{1}{J}\sum_{k=1}^S \left(\frac{u_k}{J}\right)^{n-1},\;\;\;\;\;d_n\equiv \sum_{k=0}^\infty d_{k,n} J^{-k}.   \eeq
We will later estimate the behavior of $d_{k,n}$ at $n\to\infty$ and high orders of $1/J$ expansion. This asymptotics will be universal and a few leading terms of it will be the same for  various  definitions of non-local  charges.

\section{$1/J$ expansion of BAE}

Let us start from reviewing the "old" method of solving \eq{BAE} in
the thermodynamical limit $J\to\infty$, $u_k\sim J\sim S$, before
sticking with the most efficient one using the Baxter equation.

As we mentioned in the introduction  the \eq{BAE} has only real
solutions, i.e. all the roots lie on the real axis. We label the
roots so that $u_{j+1}>u_j$. Suppose there exists a smooth function
$X(x)$ parametrizing the Bethe roots
\beq u_k=J X(k/J),\;\;\;\;\;\vrho(X(x))\equiv\frac{1}{X'(x)}\simeq\la{defvro}
\frac{1}{u_{k+1}-u_k}. \eeq
For large $S$  the function $\vrho(x)$ has a meaning of density of Bethe roots.
As follows from definition (\ref{defvro}) its normalization is
\beq   \int dx\vrho(x)=\alpha   \eeq
with $\alpha =S/J$.
Taking $\log$ of both parts of \eq{BAE} we have\footnote{Note that $\frac{i}{2}\log\frac{x+i}{x-i}
=\arctan(x)-\frac{\pi}{2}{\rm sign}(x)$ for standard definition of the $\log$ i.e.
$\log(x)=-\log(x-i0)$ for $x<0$.}
\beq 2\pi i m_j+J\log\frac{u_j-i/2}{u_j+i/2}= -{\sum_{k=1}^{\;\;\;S\;\;\textbf{,}}}
2i\arctan(u_j-u_k), \eeq
where $m_j$ are strictly ordered integers $m_{j+1}>m_j$.
In terms of  the logarithm we can write
\beq 2\pi i n_j+J\log\frac{u_j-i/2}{u_j+i/2}= {\sum_{k=1}^{\;\;\;S\;\;\textbf{,}}}
\log\frac{u_j-u_k+i}{u_j-u_k-i}, \la{logBAE}\eeq
where $n_j=m_j-j+\frac{S+1}{2}$ are non-decreesing integers
\footnote{ This fact is obvious when $S \ll J$, for $J\to\infty$, when we can neglect the r.h.s. of  \eq{logBAE}. We believe  that this is true also in general, although this fact is irrelevant for the rest of the paper}.

 Now we have instead of ``fermionic" $m_j$'s the ``bosonic"
$n_j$'s which are simply ordered $n_{j+1}\ge n_j$ and different
Bethe roots can have the same magnon numbers $n_j=n_{j+1}=\cdots$.
In the thermodynamical limit we can rewrite \eq{logBAE} assuming $k$ to be far from the edges, as follows
(see also Appendix A)
\beqa \label{RHOEXP}
&&{\sum_j}'i\log\left(\frac{u_j-u_k+i}{u_j-u_k-i}\right)
\!=\!-\!2{\sum_j}'\frac{1}{u_j-u_k}\!+\!
\frac{2}{3}{\sum_j}'\frac{1}{(u_j-u_k)^3}\!-\!
\frac{2}{5}{\sum_j}'\frac{1}{(u_j-u_k)^5}\!+\!
\frac{2}{7}{\sum_j}'\frac{1}{(u_j-u_k)^7}\\\nn
&&+\frac{\pi\vrho'[\coth(\pi\vrho)]_6}{J}
-\frac{1}{12
J^3}\left((\pi\vrho')^3
\left[\frac{\coth(\pi\vrho)}{\sinh^2(\pi\vrho)}\right]_2
-2\pi^2\vrho'\vrho''
\left[\frac{1}{\sinh(\pi\vrho)}\right]_3
+\pi\vrho^{(3)}[\coth(\pi\vrho)]_4\right)
+\O\left(\frac{1}{J^{5}}\right),
\eeqa
were we introduce the notation defined by $[f(\vrho)]_n\equiv f(\vrho)-\sum_{i=0}^{n-1} f^{(i)}(0)\frac{\vrho^i}{i!}$
for the  functions regular at zero. For singular functions  the Taylor series should be substituted by the Laurent series so that
$[f(\vrho)]_n$ is zero for $\vrho=0$ and has first $n-1$ zero derivatives at this point.
 The terms in the first
line represent the naive expansion of the l.h.s. in $1/(u_j-u_k)$.
It works well for the terms in the sum with $u_j\gg u_k$. The terms
in the second line describe the anomalous contribution at $u_j\sim
u_k$, for close roots with $i\sim j$. In this case we can expand
\beq   u_j-u_k= \frac{j-k}{\vrho(u_j/J)}+\O(1/J)   \eeq
and calculate the corresponding converging sum giving the terms in
the second line. This anomaly was noticed in the Bethe ansatz
context in \cite{Beisert:2005di} although this phenomenon was known  since
long in the large N matrix integrals or similar  character
expansions \cite{Matytsin:1993iq,Daul:1993xz}. It was proven in \cite{Beisert:2005di} to
happily cancel in the main order of $1/J$, even for a more
complicated nested Bethe ansatz.

In our case when  {$J\rightarrow\infty$} it is obvious from \eqs{RHOEXP}{logBAE} that the anomaly does not
contribute to the main order and the Bethe ansatz equation becomes a singular integral
equation \cite{Sutherland,Beisert:2003xu}
\beq   2\pi n_k-\frac{1}{x} =2\int_{C_{tot}} \frac{dy\
\vrho_0(y)}{x-y},\qquad x\in C_k,\ k=1,\cdots,K.\eeq

Now we introduce the resolvent and the quasi-momentum:
\beq\la{DEFPX} G(x)\equiv\sum_{j}\frac{1}{x J-u_j},\;\;\;\;\;
p(x)\equiv G(x)+\frac{1}{2x}
\eeq
as well as the standard definition of  density
\footnote{These two densities $\rho(x)$ and $\vrho(x)$ are clearly diferent.
By definition $\vrho(x)$ is a smooth function, whereas $\rho(x)$ is a sum of $\delta$-functions.
However their $1/J$ expansion is the same at least for the first two orders.
It is also important to note that to obtain the $1/J$ expansion of $\rho(x)$
one should first expand the rhs of \eq{defrho}
in powers of $1/J$ and then take the limit $\epsilon\rightarrow 0$.
}
\beq
\rho(x)=\frac{1}{2\pi i}\left(p(x-i\epsilon)-p(x+i\epsilon)\right)\la{defrho}
\eeq
and rewrite it in terms of the following Riemann-Hilbert problem
\beq\label{RHP}  2\sp(x)=2\pi n_k,\qquad x\in C_k,\quad n_k\in {\bf
Z}, \eeq
where $\sp(x)=\half \[p(x+i0)+p(x-i0)\]$ denotes the symmetric
(real) part of the quasi-momentum on a cut $C_k$, $k=1,\cdots,K$.

\begin{figure}[t]
\centerline{\epsfxsize=0.6 \textwidth  \epsfbox{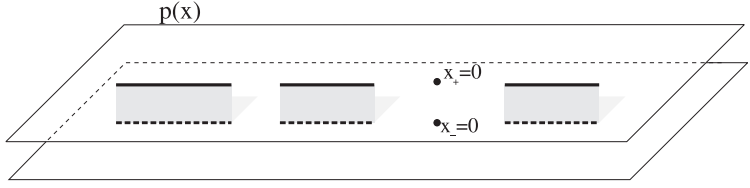}}
\caption{\label{fig1}\textit{Hyperelliptic Riemann surface}}
\end{figure}

To solve it \cite{Kazakov:2004nh}  we notice that $p(x)$ is a function on a
hyperelliptic Riemann surface with two sheets related by  $K$ cuts along the real axis, as
shown on the fig.\ref{fig1}.
 It has a known
behavior $p(x)\simeq P_0+\O(x)$ at $x=0$, and $p(x)\sim
{S/J+1/2\over x}+\O(1/x^2)$ at $x\to\infty$ on the first sheet. There are no other singularities. However, as we see from
\eq{RHP} $p(x)$  is not a single valued function but its
derivative is. This information is enough to fix $p'(x)$ as a
single-valued function  on hyperelliptic Riemann surface $f^2 =
\prod_{j=1}^{2K} (x-{\rm x}_j)$
\beq\label{QMPRIM} p'(x) = {1\over f(x)} \sum_{k=-1}^{K-1}a_k
x^{k-1}. \eeq
The single-valuedness imposes $\oint_{A_l}dp = 0$,
$l=1,\ldots,K-1$, where $A_l$, $l=1,\cdots,K$ are the $A$-cycles each
surrounding a cut $C_l$.

The BAE  \eq{RHP} become the  integer $B$-period conditions:
\beq \oint_{B_j}dp = 2\pi n_j\qquad  n_j=1,\ldots, {K}, \eeq
where the cycle $B_{j}$ starts at
$x=\infty_+$ on the upper sheet, goes through the cut $C_J$ to the
lower sheet and ends up at its infinity $x=\infty_-$.

 From \eq{BAE} we have the quasi-periodicity condition for the total
momentum expressed by \eq{QPC}. In some applications, such as the
above mentioned spectrum of the integrable matrix of anomalous
dimensions in ${\cal N}=4$ SYM theory, we select  only purely
periodic Bethe states (\ref{periodic}).
This information, together with the filling fractions $\alpha_k=S_k/J$,
$k=1,\cdots,K$, such that  $\sum_{k=1}^K \alpha_k=S/J$, or
\beq  \alpha_k=\oint_{A_k} dx\ p(x)  \eeq
fixes completely a solution.
\begin{figure}[t]
\centerline{ \epsfxsize=0.6\textwidth \epsfbox{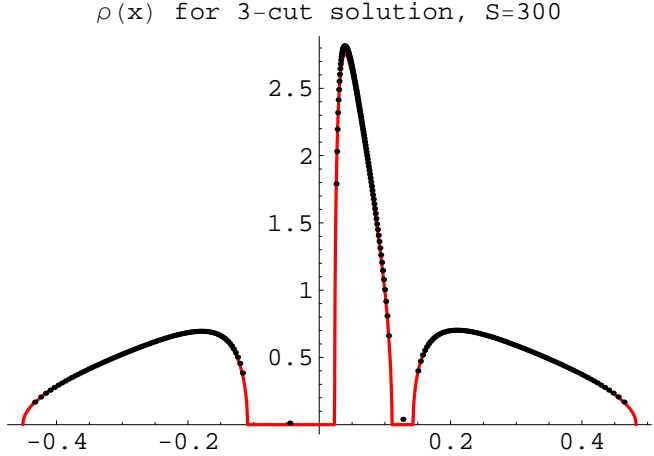}}
\caption{\label{fig:Ai}\textit{Density of  roots. The dots correspond to  numerical 3-cut
solution with total number of Bethe roots $S=300$ and equal
fractions $\alpha_i=1/6$,  and $n_i=\{-1,3,1\}$. They are fixed from
the numerical values of the roots by the \eq{defvro}.  Solid line is the density at
$J=\infty$ computed analytically from the corresponding
hyper-elliptic curve. $x$ coordinates of the dots are $\frac{u_{j}+u_{j+1}}{2J}$
so that the solitary points in the middle of empty cuts are artifacts of this definition.}}
\end{figure}
For the energy we obtain in this limit
\beq\label{ENZERO}  E_0=-G_0'(0).  \eeq
A particular 3-cut solution is demonstrated on the fig.2 and is
compared with the numerical solution of exact Bethe ansatz equation. We see that
already for 300 roots and $J=600$ the description in terms of the
algebraic curve becomes excellent.

\section{Large $J$ limit and $1/J$-corrections from Baxter equation}

Eq. (\ref{BAE}) can be also obtained as the condition that the
transfer matrix eigenvalue defining a Bethe state (see for example \cite{Krichever:1996qd})
\beq\label{BAXTER}
T(u)=W(u+i/2)\frac{Q(u+i)}{Q(u)}+
W(u-i/2)\frac{Q(u-i)}{Q(u)},
\eeq
where $Q(u)=\prod_{k=1}^S (u-u_k)$, $W(u)=u^J$. $T(u)$ is a polynomial of
degree $J$, which is clear from the very  construction of a Bethe
state in the algebraic Bethe ansatz approach \cite{Faddeev}.
The Bethe equations (\ref{BAE}) follow immediately from \eq{BAXTER}  assuming analyticity (polynomiality)
of $T(u)$.

Introduce the notations: $x=u/J$,
$\Phi(x)=\frac{1}{J}\sum_{k=1}^S\log(x-x_k)$, $V(x)=\log x$,
$2t(x)=T(Jx)/(Jx)^J$ and rewrite (\ref{BAXTER}) as
\beq\label{BAXTEXP} 2t(x)=\exp J\left[\Phi\left(x+\frac{i}{J}\right)
-\Phi(x)+V\left(x+\frac{i}{2J}\right)-V(x)\right] +{\rm c.c.}. \eeq
Defining the quasi-momentum (now at all orders in $1/J$)
\beq\label{DEFQM} p(x)\equiv\Phi'+V'/2  \eeq
and  expanding the Baxter equation in $1/J$ we get
\beqa
\label{BAXEX} t(x)&=&\cos
p(x)\left[1-\frac{1}{J}\left(\frac{p'(x)}{2}-\frac{V''(x)}{8}\right)
+\frac{1}{2J^2}\left(\frac{p'(x)}{2}-\frac{V''(x)}{8}\right)^2\right]\\
&+&\nn\frac{1}{J^2}\sin p(x)\left(\frac{p''(x)}{6}-\frac{V^{(3)}(x)}{16}\right)+
O\left(\frac{1}{J^3}\right).
\eeqa
To find the $1/J$ corrections to
the quasi-momentum, we will expand $p(x)=p_0(x)+\frac{1}{J}p_1(x)+
\frac{1}{J^2}p_2(x)+O(1/J^3)$, $t(x)=t_0(x)+\frac{1}{J}t_1(x)+
\frac{1}{J^2}t_2(x)+O(1/J^3)$ and plug it into the last equation. We
assume that the coefficients of expansion
$t_0(x),t_1(x),t_2(x),\ldots$ are the entire functions on the plane
$x$ with no cuts,  having only an essential singularity at $x=0$.

The quasi-periodicity property of the total momentum reads up to 3 first
orders as follows
\beq\la{QMEX}   P_{tot}=-\sum_j\frac{1}{u_j}+\sum_j\frac{1}{12u^3_j}+\O(1/J^4)=2\pi k/J   \eeq
and in the purely periodic case we select only the states with $k=m J$, with integer $m$.

\subsection{Zero order from Baxter equation}

Let us restore from the Baxter
equation the zero order result of the previous section. In the  zero order approximation we get from \eq{BAXEX}
\beq\label{ZEROP}  \cos p_0(x)=t_0(x)  \eeq
or
\beq\label{ZEROSOL}
p'_0(x)=\frac{2t_0'(x)}{\sqrt{1-t_0^2}}.  \eeq
As  is usual in the finite gap method \cite{Novikov:1984id}, we expect
that since $t_0(x)$ is an entire functions all the branch cuts of
\eq{QMPRIM} come from the square root in denominator, after the Bethe roots condense to a set
$C_1,\cdots,C_K$ of dense supports in the $J\to\infty$ limit. It is
easy to see from the definition \eq{DEFQM} that $p_0(x)= \frac{\a+1/2}{x}+\O(1/x^2)$
when $x\to\infty$ and $p_0(x)\sim 2\pi \frac{k}{J}+ \O(x)$ when $x\to 0$. Consequently,
we reproduced the general solution (\ref{QMPRIM}-\ref{ENZERO}) of
\cite{Kazakov:2004nh}.

\subsection{$1/J$ correction from Baxter equation}
To find  the next, $1/J$ approximation to the density of roots  and
to the energy we deduce from \eq{BAXEX}
\beq\label{PONE}  p_1=(-p_0'/2+V''/8)\cot
p_0-\frac{t_1}{\sin p_0}, \eeq
where $t_1(x)$ is an entire  function on the plane as was mentioned
before.  We know about $p_0(x)$ that
\beq p^+_0=\pi n_j-\pi
i\rho_0,\;\;\;\;\;p^-_0=\pi n_j+\pi i\rho_0,\;\;\;\;\;\sin
p_0^+=-\sin p_0^-\la{pppm} \eeq
and thus we have for the real and imaginary parts of $p_0(x)$ on the cuts
\beq \pi i\rho_1=\left(\frac{V''}{8}t_0-t_1\right)\frac{1}{\sin
p_0^-},\qquad \sp_1=-p_0' \cot p_0/2.   \la{rez1}\eeq
 We will solve these equations below and restore the explicit $p_1$.

Since $t_1(x)$ is a regular function on the cuts (\cite{Korchemsky:1995be})
\beq  \sp_1=-\frac{1}{2}\pi\rho'\coth\pi\rho=-{p_0^\pm}'\cot
p_0^\pm/2 \la{maineq}.   \eeq
Moreover we see  from \eq{QMEX} that
\beq
p_1(0)=0\la{p1x0}
\eeq
and for large $x$ $p_1(x)$ should decreases as $\O(1/x^2)$.

We can write from (\ref{maineq})   the general solution of this
Riemann-Hilbert problem
\beq\la{p1res} p_1(x)=\frac{x}{4\pi i f(x)}\oint_{{\cal
C}}\frac{f(y)p_0'(y)\cot p_0(y)}{y(y-x)}dy+\sum_{j=1}^{K-2}\frac{a_j
x^j}{f(x)}\;, \eeq
where $f^2(x)=\prod_{j=1}^{2K}(x-x_j)$ and the contour encircles all
cuts $C_k$ (but no other singularities). The first term in the
r.h.s. represents the Cauchy integral restoring the function from
its real part on the cuts and having a zero at the origin (the value
of the quasi-momentum $p(x)$ at $x=0,\infty$ was already fixed for
$p_0$) whereas the second one is purely imaginary on the cuts, with
the polynomial in the numerator chosen in such a way that it does
not spoil the behavior of $p(x)$ at $x=0,\infty$.\footnote{
We could also add terms $\frac{1}{f^3}$, $\frac{1}{f^5}$, $\dots$ but
they are too singular at the branch points as we shell see in the
next section.}

Thus for $K<3$ the solution is unique. In particular, for $K=1$ we
restore from here the 1-cut solution of
\cite{Beisert:2005mq}. For $K\geq 3$ we have to fix $K-2$
parameters $a_j$. To do this we can use $K$ additional conditions
ensuring the right fractions $\alpha_j$ of the roots already chosen
for $p_0$:
\beq \oint_{C_l}p_1(x)
dx=0,\;\;\;\;\;l=1,\dots,K,\la{cond}
\eeq
in fact only $K-2$ of them are linear independent (since we have
already fixed the total filling fraction by the asymptotic
properties of \eq{p1res} at $x=\infty$: $p_1(x)=O(1/x^2)$.
\Eq{p1x0} also restricts some linear combination of the conditions
(\ref{cond})). Hence we completely fixed all
parameters of our $K$-cut solution for the $1/J$ correction $p_1$
knowing the zero order solution (algebraic curve) for $p_0$.

It is also  useful to rewrite \eq{p1res} in the following way
\beq p_1(x)=\frac{Q(x)}{4\pi i f(x)}\oint_{{\cal
C}}\frac{f(y)p_0'(y)\cot p_0(y)}{Q(y)(y-x)}dy\;, \eeq
where $Q(x)=\sum_{k=1}^{K-2}\tilde a_j x^j$ and the contour of
integration  encircles  all the cuts. Again, $\tilde a_j$ are fixed
by \eq{cond}. Equivalence of this formula to \eq{p1res} can be seen
from the coincidence of their analytical properties: blowing up the
contour in the  contour integrals of any of these representations we obtain
\footnote{In fact one should take into account an infinite number
of residues at $y=0$, see \cite{Beisert:2005mq}. A more regular
 procedure is to express  $\cot$
as a sum and then do the integration.}
\beq p_1(x)=-p_0'(x)\cot p_0(x)/2+\frac{1}{f(x)} \sum_n
\frac{c_n}{x_n-x} +\sum_{j=1}^{K-2}\frac{a_j x^j}{f(x)},  \eeq
where the first term comes from the residue at $y=x$. Note that only
this term contributes to the r.h.s of  \eq{maineq} thus showing that
we satisfied this equation. The second term exactly cancels all the
poles of the first term, so that $p_1$ is regular everywhere except
the cuts. It is a meromorphic function in the $x$ plane, having no
cuts. The last term  reflects the freedom of adding $K-2$
coefficients before fixing them by conditions \eq{cond}.

\subsection{Equations for $1/J^2$ corrections   from Baxter relation}

Expanding \eq{BAXEX} up to $1/J^2$ we obtain

\beq p_2=-\frac{1}{2}\d_x[\cot(p_0)I]-\frac{1}{8 x^3}-\frac{\tilde
t_2}{2\sin(p_0)} \la{Baxterp2}, \eeq
where
\beq I=-\frac{\tilde t_1}{\sin(p_0)}=p_1+\frac{p_0'}{2}\cot p_0.
\eeq
 We introduced here  the notations
\beq
\tilde t_1=t_1+\frac{\cos p_0}{8x^2},\;\;\;\;\;
\tilde t_2=t_2-\frac{\cos p_0}{128x^4}+\frac{\tilde t_1}{8x^2}-\frac{\cos(2p_0)+5}{24\sin p_0}p_0''
+\frac{\cos p_0}{8\sin^2 p_0}\left(3(p_0')^2+4{\tilde t}_1^2\right)
\eeq
so that $\tilde t_1$ and $\tilde t_2$ are single valued functions.

Note that  above the  cut $I^+=\pi i\rho_1$. We will find the
explicit solution of these equations later, but we will need for
that some results of the next section where we study the behavior of
$p(x)$ near the branch points.

\section{ Double scaling solution near the branch point}

\begin{figure}[t]
\centerline{ \epsfxsize=1\textwidth \epsfbox{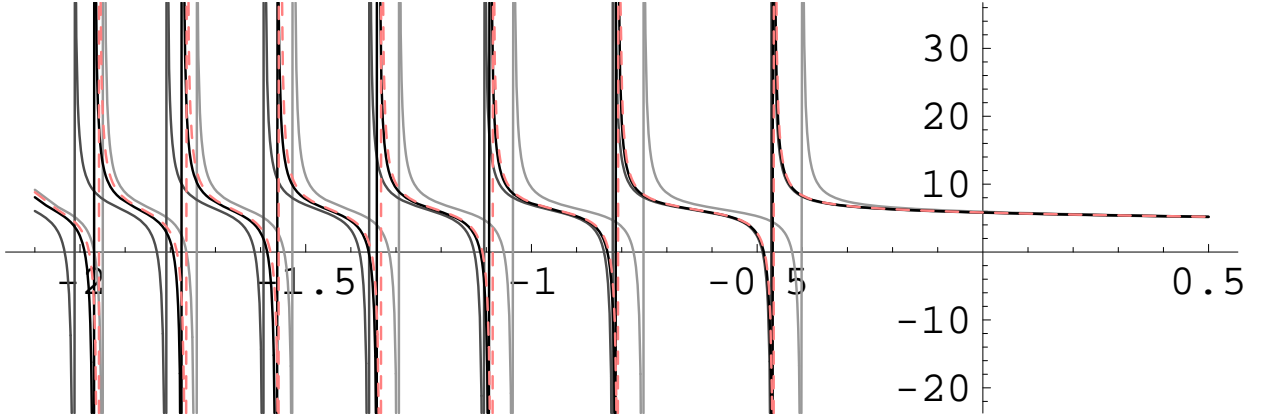}}
\caption{\textit{Quasi-momentum near branch point as a function of the
scaling variable $v$ for $S=200$. The poles corresponds to the positions of
Bethe roots $u_i$. Red dashed line - "exact" numerical value, light
grey - zero order approximation given by Airy function ${\rm
Ai}(a^{1/3}x)$, grey - first order and black - second order
approximation.}  \label{fig:Ai}}
\end{figure}

As we stated above the branch point singularities come only from the
square roots of the denominator of \eq{ZEROSOL}. We define an exact
branch point as a point $x_*$ where $t(x_*)=\pm 1$
. If we approach one
of the branch points $x\to x_*$ we can expand
\beq\la{EXPT} t(x)\simeq
\pm[1-a(x-x_*)/2-b(x-x_*)^2/2].
\eeq
  Note that $x_*, a, b$ themselves  depend
on $J$. We assume that they have a regular
expansion in $1/J$ (justified by  self-consistency of  further
calculations and by  numerics) and define $x_*=x_0+x_1/J+\dots$. We
call $x_0$ a classical branch point and $x_1/J$ a branch point
displacement.

 Denoting $v=(x-x_*)J^{2/3}$
which will be our double scaling variable $v\sim 1$, we get  from \eq{BAXTER} up to
$1/J^2$ terms
\beq\label{BAXTL} \pm 2\left(1- \frac{a
v}{2J^{2/3}}-\frac{bv^2}{2J^{4/3}}\right)Q(u)=
Q(u+i)\frac{W(u+i/2)}{W(u)}+Q(u-i)\frac{W(u-i/2)}{W(u)}.
 \eeq
In terms of a new function
$q(v)=e^{-n\pi v J^{1/3}}e^{\frac{v J^{1/3}}{2x_*}}Q(x_*J+v
J^{1/3})$, where $n$ is such that $t(x^*)=e^{i \pi n}$, and after expansion in $1/J$ the last eq. takes the form
\beq\la{dseq} q''-a v q=\frac{1}{J^{1/3}}\frac{4vq'+q}{4x_*^2}
+\frac{1}{J^{2/3}}\left[ \frac{1}{12}q^{(4)}(v)-
\frac{v^2q(v)}{4}\left(\frac{1}{x_*^4}-4b\right)\right]
+\O\left(\frac{1}{J}\right) .\eeq
In fact, this equation can be easily
solved in terms of $q_0$
\beq\la{qres} q\propto\left[1+\frac{v^2}{4 x_*^2
J^{1/3}}+\frac{1}{J^{2/3}}
\left(\frac{v^4}{32x_*^4}-\frac{3b-a^2}{15a}v\right) \right]q_0\!\!
\left(v-\frac{1}{4ax_*^2J^{1/3}}+\frac{a^2+12 b}{60 a
J^{2/3}}v^2\right), \eeq
 where $q_0(v)={\rm Ai}(a^{1/3}v)$ (the Airy
function). The second solution of the \eq{dseq}, ${\rm Bi}(a^{1/3}v)$  has
a  wrong asymptotic as we will see. The sign $\propto$ means that
the solution is defined up to a constant multiplier but this
unknown multiplier doesn't affect the quasi-momentum. Now we can express the
quasi-momentum only through our scaling function $q(v)$
\beq\la{AiryF}
p\left(x_*+\frac{v}{J^{2/3}}\right)=\frac{\d_v
q(v,J)}{q(v,J)J^{1/3}}+ \pi n +\frac{1}{2x_*} \left(\frac{1}{1+\frac{v}{x_*
J^{2/3}}}-1\right).\eeq
The first two terms in the r.h.s., if we substitute $q(v)\to q_0(v)$, represent the principal contribution to the double scaling limit near the edge, valid up to the corrections of the order $1/J^{2/3}$. We  see from the definition (\ref{DEFPX}) that the   zeros of $q(v)$ are nothing
but the positions $u_i$ of  Bethe roots. Thus we know  these
positions
  with a precision $1/J^{2/3}$ (see fig.3).

 The large $v$ asymptotic will
be very helpful in fixing some unknown constant in the $1/J^2$
corrections given in the next section
\beqa &&p(\!x_*\!+\!v J^{-2/3}\!)\!=\!\pi
n\!+\! \frac{1}{J^{1/3}}\!\!\left(\!-\!\underbrace{\sqrt{a
v}}_1\!-\!\!\underbrace{\frac{1}{4 v}}_{1/J}\!\!+\!
\underbrace{\frac{5}{32v^2\sqrt{av}}}_{1/J^2} +\dots\!\right)\!\!
+\!\frac{1}{J^{2/3}}\!\!
\left(\!\underbrace{\frac{1}{8x_*^2\sqrt{av}}}_{1/J}
\!-\!\underbrace{\frac{1}{16 a x_*^2 v^2}}_{1/J^2}\!+\!\dots\!\right)\!\!+\dots
\la{p0exp}, \eeqa
where the cut corresponds to
negative $v$ for $a>0$. Introducing the notation $y=vJ^{-2/3}$ and
rearranging the terms by the powers $1/J$ we have
\beqa \la{seriesp} p(x_*+y)&=&\pi
n+\left[-\sqrt{ay}- \frac{(a^2+12b)y^{3/2}}{24\sqrt{a}}+\dots\right]
+\frac{1}{J}\left[-\frac{1}{4y}+
\frac{1}{8x_*^2\sqrt{ay}}+\frac{a^2-4b}{16a}
+\dots\right]\\
\nn&&+\frac{1}{J^2}\left[\frac{5}{32y^2\sqrt{ay}}
-\frac{1}{16 a y^2 x_*^2} +\frac{6-x_*^4(a^2+12
b)}{768x_*^4(ay)^{3/2}}+\dots\right]+\dots .
\eeqa
Doing this re-expansion   we assume that $J^{-1}\ll y \ll 1$, trying
to sew together the double scaling region with the $1/J$ corrections
to the thermodynamical limit. This  procedure is similar to the one used in higher orders of the WKB approximation in the usual one dimensional quantum mechanics (see for example \cite{Bleher}).

 To compare with $p_0,\;p_1$ and $p_2$
we have to re-expand around $x_0$
\beq
p(x_0+y)=p(x_*+y)+\frac{x_1}{J}
\frac{\sqrt{a}}{2\sqrt{y}}+
\frac{1}{J^2}\left[-\frac{x_1}{4y^2}+
\frac{x_1}{16x_0^2y\sqrt{a
y}}+\frac{\sqrt{a}x_1^2}{8y\sqrt{y}}\right]
\eeq
or, introducing notation
\beq  x_1=\frac{2A}{\sqrt{a}}-\frac{1}{4x_0^2a}  \eeq
 we get
\beqa\la{p0branchP}
 p(x_0+y)&=&\pi
n+\left[-\sqrt{ay}-
\frac{(a^2+12b)y^{2}}{24\sqrt{ay}}+\dots\right]
+\frac{1}{J}\left[-\frac{1}{4y}+
\frac{A}{\sqrt{y}}+\frac{a^2-4b}{16a}
+\dots\right]\\
\nn&&+\frac{1}{J^2}\left[\frac{5}{32y^{2}\sqrt{ay}}
-\frac{A}{2 \sqrt{a} y^2} +
\left(\frac{A^2}{2y\sqrt{a y}}-\frac{b}{64(ay)^{3/2}}-\frac{\sqrt{a y}}{768y^2}\right)+\dots\right]+\dots.
\eeqa
Near the left branch point (i.e for $a<0$ and $y<0$) we have
\beqa\la{p0branchJ}
 p(x_0+y)&=&\pi
n+\left[\sqrt{ay}+
\frac{(a^2+12b)y^{2}}{24\sqrt{ay}}+\dots\right]
+\frac{1}{J}\left[-\frac{1}{4y}-
\frac{A}{\sqrt{-y}}+\frac{a^2-4b}{16a}
+\dots\right]\\
\nn&&+\frac{1}{J^2}\left[-\frac{5}{32y^{2}\sqrt{ay}} +\frac{A}{2
\sqrt{-a} y^2} + \left(\frac{A^2}{2y\sqrt{a
y}}+\frac{b}{64(ay)^{3/2}}+\frac{\sqrt{a
y}}{768y^2}\right)+\dots\right]+\dots. \eeqa
 Now we can compare it
with our  results of the previous sections and fix $a,b$ and $x_1$.

Let us note that similar Airy type oscillations were observed in the papers on random matrices where this behavior occurs near  an endpoint of a distribution of eigenvalues \cite{BrezinKazakov}. This shows the intrinsic similarity of these two seemingly different problems. The large $J$ limit for the spin chain is rather similar
to the large size $N$ limit in random matrices. We will see this analogy even clearer in the section  7 where we will calculate the asymptotics of conserved charges using the results of this section.

\subsection{Comparison with $p_0$ and $p_1$}

It is instructive  to establish the relations between $a,b,A$ and the
parameters of the algebraic curve which completely defines, as we know,
$p=p_0+\frac{1}{J}p_1+\O\(\frac{1}{J^2}\)$ up to the first two orders.

For that we use the expansion (\ref{EXPT}) defining $a,b$ and find from
\eq{ZEROP}
 for $y>0$
\beq p_0(x_0+y)=\pi n+\arccos t_0\simeq\pi n-\sqrt{ay}-\frac{a^2+12
b}{24\sqrt{a}}y^{3/2}+\O(y^{5/2}), \eeq
in agreement with \eqs{p0branchP}{p0branchJ}.
We can fix $a$ and $b$ up to $\O\(1/J\)$ corrections from here through
the parameters of the solution for $p_0$ given by  \eq{QMPRIM}.

To calculate $a$ and $b$ up to $\O\(1/J\)$ and to fix $A$,
we use the expansion \eq{EXPT} with \eq{PONE}.
Note that we have the minus sign in front of $\sqrt{ay}$ which
ensures the positivity of the density (\ref{defrho}) on the cut (i.e. for $y<0$
and $a>0$)
$\rho(y)\simeq \sqrt{a(-y)}/\pi$. If we had $\rm Bi$ instead of $\rm Ai$
the sign would be plus and the density would be negative.

Now we compare  this near-cut behaviour to $p_1$. Consider the regular part first
\beq \sp_1=-\frac{1}{2}p'_0\cot
p_0\simeq -\frac{1}{4y}+\frac{a^2-4b}{16a}+\O(y), \eeq
which is  in full agreement with \eq{seriesp}. From \eq{p1res} we
see that
\beq p_1(x_0+y)-\sp_1(x_0+y)\simeq
\frac{A}{\sqrt{y}}+\O\left(\frac{1}{y^{3/2}}\right), \eeq
 where $A$
can be written explicitly,
again using the parameters of $p_0$ given by  \eq{QMPRIM}.

For the  example of  one-cut solution see  \eq{A1cut}.

\section{ General solution for $p_2$ and $E_2$ }

Now we have enough of information to construct $p_2$ in the
most general situation of an arbitrary number of cuts.

We start from a formula which immediately follows from
\eq{Baxterp2}
\beq\la{sp2} \sp_2=-\frac{1}{2}\d_x\left[\cot(p_0)
\left(p_1+\frac{p_0'}{2}\cot p_0\right)\right]-\frac{1}{8x^3}, \eeq
where $p_1$ is given by \eq{p1res}. The behaviors near zero and at
infinity are the following. Since from \eq{QPC} and
\eq{QMEX} it follows that
$G(0)-\frac{1}{24J^2}G''(0)=2\pi k/J+\O(\frac{1}{J^4})$  we can conclude that
\beq\la{condp20} p_2(0)=\frac{1}{24}G''_0(0). \eeq
For large $x$ we have again
\beq\la{condp2infty} p_2(x)=O\left(1/x^2\right). \eeq
Repeating the arguments of the previous subsection we have
\beq\la{p2solution} p_2(x)=\frac{x}{4\pi i f(x)}\oint_{\cal C}
\frac{f(y)}{y(y-x)}\left(\frac{1}{4y^3}+
\d_y\left[\cot(p_0)p_1\right]\right) +\sum_{j=0}^{5K-1}\frac{c_j
x^j}{f^5(x)}, \eeq
where the path ${\cal C}$ is defined as in \eq{p1res}. Again the first term guarantees that $p_2$
satisfies the \eq{sp2}. We drop out the $p_0'\coth p_0$ for simplicity.
We can do this since together with $f(y)$ it forms a single-valued
function without cuts and the integral is given by the poles inside of
the path of integration. In fact there are only poles at each branch point so that the result can be absorbed into the second term in
\eq{p2solution}.

So far the second term in \eq{p2solution} was  restricted only by the
conditions (\ref{condp20}) and (\ref{condp2infty}). Of cause this
does not explain why we should restrict ourselves by the fifth power
of $f(x)$ in denominator. A natural explanation comes from the
known behaviour near the branch points
(\ref{p0branchP},\ref{p0branchJ}) from where we can see that
\beq\la{p2PL}
p_2(x^i_0+y)=\left\{
\bea{c}
\frac{5}{32y^{2}\sqrt{a_iy}}
-\frac{A_i}{2 \sqrt{a_i} y^2} +
\left(\frac{A_i^2}{2y\sqrt{a_i y}}
-\frac{b_i}{64(a_iy)^{3/2}}-\frac{\sqrt{a_i y}}{768y^2}\right)
+\O\left(\frac{1}{y}\right),\;\;\;\;\;a_i>0,\;y>0\\
-\frac{5}{32y^{2}\sqrt{a_iy}}
+\frac{A_i}{2 \sqrt{-a_i} y^2} +
\left(\frac{A_i^2}{2y\sqrt{a_i y}}+\frac{b_i}{64(a_iy)^{3/2}}
+\frac{\sqrt{a_i y}}{768y^2}\right)
+\O\left(\frac{1}{y}\right),\;\;\;\;\;a_i<0,\;y<0
\eea
\right.,
\eeq
where all $6K$ constants $a_i,\;b_,\;A_i$ for $i=1,\dots,2K$ are
known
 since they can be determined from the
near branch point behaviour of $p_0$ and $p_1$
(\ref{p0branchP},\ref{p0branchJ}). $a_i$ and $b_i$ follow from $p_0$
\beq
p_0(x^i_0+y)=\left\{
\bea{c}
-\sqrt{a_iy}-
\frac{(a_i^2+12b_i)y^{2}}{24\sqrt{a_iy}}
+\O\left(y^{5/2}\right),\;\;\;\;\;a_i>0,\;y>0\\

\sqrt{a_iy}+
\frac{(a_i^2+12b_i)y^{2}}{24\sqrt{a_iy}}
+\O\left(y^{5/2}\right),\;\;\;\;\;a_i<0,\;y<0
\eea
\right.
\eeq
and $A_i$ comes from $p_1$
\beq
p_1(x^i_0+y)=\left\{
\bea{c}
-\frac{1}{4y}+\frac{A_i}{\sqrt{y}}
+\O\left(y^{0}\right),\;\;\;\;\;a_i>0,\;y>0\\
-\frac{1}{4y}-\frac{A_i}{\sqrt{-y}}
+\O\left(y^{0}\right),\;\;\;\;\;a_i<0,\;y<0 \eea \right.. \eeq
 In
fact \eq{p2PL} gives only two nontrivial conditions for each branch
point which are the coefficient before the half-integer power of $y$ so
that we have $4K$ conditions. The extra $K$ conditions come from zero
$A$-period constraints signifying the absence of corrections to the filling fractions $\alpha_i$.
\beq\la{ACYC} \oint_{{\cal C}_l}p_2(x) dx=0,\;\;\;\;\;l=1,\dots,K.
\eeq

To reduce the number of unknown constants consider a branch point $x_0$.
We can see that for small $y=x-x_0$ (for simplicity we assume that the
cut is on the left i.e. $a_i>0$)
\beq
I_1\equiv\frac{x}{4\pi i f(x)}\oint_{\cal C}
\frac{f(z)}{z(z-x)}\left(\frac{1}{4z^3}+\d_z(p_1\cot\!p_0)\right)
=\frac{3}{16y^2\sqrt{ay}}-\frac{A}{2\sqrt{a}y^2}
+\frac{1}{y^{3/2}}\left(\frac{b}{32a^{3/2}}
-\frac{5\sqrt{a}}{128}\right)+\O\left(\frac{1}{y}\right).
\eeq
Introduce the following integral
\beq
I_2\equiv\frac{x}{4\pi i f(x)}\oint_{\cal C}
\frac{f(z)}{z(z-x)}\left((p_1+p_0'\cot\! p_0)p_1\cot\! p_0-\frac{p_0''}{12}\right)=
-\frac{1}{32y^2\sqrt{ay}}
+\frac{1}{y^{3/2}}\left(\frac{A^2}{2\sqrt{a}}-\frac{3b}{64a^{3/2}}
+\frac{29\sqrt{a}}{768}\right)+\O\left(\frac{1}{y}\right)
\eeq
so that $I_1+I_2$ reproduces the right series expansion near the branch
points given by \eq{p0exp} and \eq{seriesp}. Moreover,
on the cuts $I_2(x+i0)+I_2(x-i0)=0$ since the function under integral is
single valued. We can simply take
\beq
p_2(x)=I_1(x)+I_2(x)+\sum_{j=0}^{K-1}\frac{\tilde c_j x^j }{f(x)}\la{p2Gx},
\eeq
where the remaining $K$ constants are fixed from  \eq{ACYC}.
Using that $p_2(0)=G''(0)/24$ we can fix one constant  $\tilde c_0=\frac{G''(0)f(0)}{24}$ before imposing the condition (\ref{ACYC}).

This is our final result for the second quantum correction to the
quasi-momentum. In the Appendix B we will specify  this result for the example of the one-cut solution where it can be made much more explicit.
\begin{figure}[t]
\centerline{ \epsfxsize=1\textwidth \epsfbox{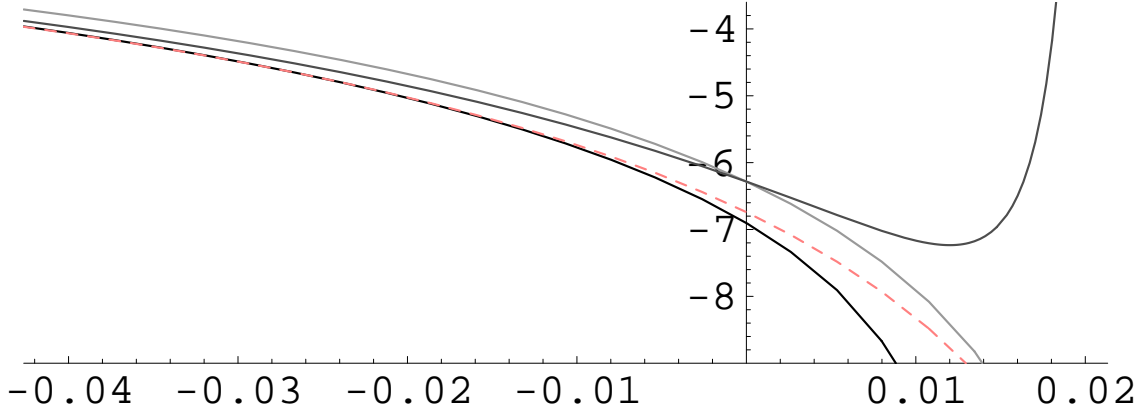}}
\caption{\label{fig:E2}\textit{Resolvent far from branch point
as a function of $x$. Red dashed line - "exact" numerical value
for one cut solution with $S=10,\;n=2,\;m=1$, light
grey - zero order approximation, grey - first order given by
\eq{p1res}
and black - second order
approximation given by \eq{p2Gx}.
Note that near branch point ($x_0=0.02$)
the approximation does not work and instead of it
we should use  the Airy function of \eq{AiryF}, like in the usual WKB near a turning point.
}}
\end{figure}

\section{Energy and higher charges}

\subsection{Energy}

To find $1/J$ corrections to the energy we represent the exact
formula \eq{ENE} as follows
\beq
E= {-}\frac{1}{J} G'(0)  +\frac{1}{ {24}J^3}
G^{(3)}(0)+\O\left(\frac{1}{J^5}\right),
\eeq
 where $G(x)$ is defined in \eq{DEFPX}. We still have to expand $G(x)=
  -\frac{1}{2x}+p_0(x)+\frac{1}{J}p_1(x)+
  \frac{1}{J^2}p_2(x)+O(1/J^3)$.

Finally, we obtain for the energy:
\beq E=\frac{1}{J}E_0(x)+\frac{1}{J^2}E_1+\frac{1}{J^3}E_2 +\O\left(\frac{1}{J^4}\right),
\eeq
where
\beqa\la{EZERO} E_0&=&
 {-}G_0'(0),\\
 \la{EONE} E_1&=& {-}p_1'(0)= {-}\frac{Q'(0)}{4\pi i f(0)}\oint_{{\cal
C}}\frac{f(y)p'(y)\cot p(y)}{Q(y)y}dy\la{E1fnl}, \eeqa
and $Q(x)=\sum_{k=1}^{K-2}b_k x^k$ is related to the last term in
(\ref{p1res}). For $E_2$ we have from
\eq{p2Gx} the following representation
\beqa\la{ETWO} E_2= \frac{G_0^{(3)}(0)}{ {24}} -p_2'(0)
&=&\frac{-1}{4\pi i f(0)}\oint\frac{f(y)}{y^2}\left(\!\frac{1}{4z^3}
+\d_z(p_1\cot p_0)-\frac{p_0''}{12}+(p_1+p_0'\cot p_0)p_1\cot p_0\!\right)\nn\\
&-&\frac{c_1}{f(0)}+\frac{G''_0(0)f'(0)}{24 f(0)}+\frac{G_0^{(3)}(0)}{24}.
\eeqa
Note that for 1-cut we should take $c_1=0$.
We can compare our results with numerical calculations, as it is
done for a few 1-cut solutions in the fig.\ref{fig:E2}

\begin{figure}[t]
\centerline{ \epsfxsize=0.6\textwidth \epsfbox{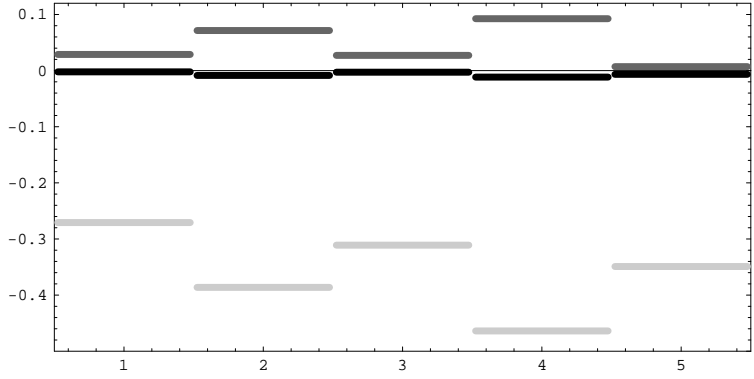}}
\caption{\label{fig:E2}\textit{Relative deviation $\delta E(S)/E(S)$ of analytical
computations of the energy $E(S)$ from its "exact" value
$E_{exact}(S)$ for the one cut distribution found numerically by
Mathematica (solid line corresponds to $\delta E(S)=0$), for a
finite number of roots $S$ and a finite length $J$ for zero order
(light gray), first order (gray) and second order (black)
approximation. Details are summarized in the table}}
\begin{center}
\begin{tabular}{|c|c|c|c|c|c|}
  \hline
   \#       & \textsf{1}    & \textsf{2}& \textsf{3}    & \textsf{4}    & \textsf{5 }   \\ \hline\hline
   $m,n$    & $1,2$         & $2,1$     & $1,3$         & $2,2$         & $1,5$         \\ \hline
   $E_0$    & $12\pi^2$     & $24\pi^2$ & $16\pi^2$     & $32\pi^2$     & $24\pi^2$     \\
   $E_1$    & $-558.4$      & $-1563$   & $-855.3$      & $-2401$       & $-1563$       \\
   $E_2$    & $1160$        & $5464$    & $1592$        &  $8982$       & $1504$        \\ \hline
   $S$      & $10$          & $40$      & $7$           & $20$          & $5$           \\
   $J$      & $20$          & $20$      & $21$          & $20$          & $25$          \\
   $E(S)$ & $4.66004$       & $8.54515$ & $5.7359$      & $10.7876$     & $7.0232$      \\
$E^{(2)}(S)$ & $4.670$      & $8.619$ & $5.752$      & $10.912$     & $7.070$      \\ \hline
\end{tabular}
\end{center}
\end{figure}

\subsection{Local charges}

In this and the next subsection we will calculate local and non-local, or global charges $Q_r$
in \textit{all} powers of $1/J$ for the large $r$ from the behavior near the relevant branch point. The idea of this  calculation  is taken from the double scaling approach in  matrix models. Namely, we can compare it to the calculation of the resolvent of eigenvalues in a gaussian unitary matrix ensemble
\beq\la{GMM}     H_N(x)= \int \frac{d^{N^2}\!\! M}{(2\pi)^{N^2}} \exp \(-\frac{N}{2}\Tr\ M^2\)\Tr(x-M)^{-1}   = \sum_{g=1}^\infty N^{2-2g} \sum_{n=0}^\infty x^{-2n-1} H_{(g,n)}\eeq
where $M$ is a hermitian matrix of large size $N$. The coefficients $H_{(g,n)}$ actually give the number of specific planar graphs: it is given by the number of surfaces  of genus $g$ which can be done from a polygon with  $2n$ edges, by the pairwise gluing of these edges.  To extract the large $n$ asymptotics of $H_{(g,n)}$ for any $g$ one can use that in the large $N$ limit the density (which is the imaginary part of the resolvent on the support of eigenvalues) is given by the Wigner's semi-circle law, and the near-edge behavior is described by the Airy functional asymptotics \cite{BrezinKazakov,MehtaBook} showing the traces of individual eigenvalues in the continuous semi-circle distribution. We will try to extract the similar asymptotics for the distribution of Bethe roots. The role of $1/N$ expansion will be played by the $1/J$ expansion, whether as the order of the $1/x$ expansion in the matrix model will be now played by the label $r$ of the charge.

We start from  expanding \eq{localDef}
\beq
Q_r=\sum_{m=0}^\infty\frac{1}{J^{r+2m-1}}\frac{(-1)^{m+1}G^{(r+2m-1)}(0)}{(2 m+1)!(r-1)!2^{2 m}}\;.\la{Qdecomp}
\eeq

As we shell see, for large $r$ only the $m=0$ term contributes. We express the
derivative as a contour integral around cuts
\beq
G^{(n)}(0)=-\frac{n!}{2\pi i}\oint_{\cal C} \frac{G(x)}{x^{n+1}}dx.
\eeq
For large $n$ only a small neighborhood of the closest to
zero branch point $x_0$ contributes due to the exponential suppression
by the $1/x^{n+1}$ factor.
Near the branch point $x_0$ we have from  \eq{p0branchP} (see also \eqs{p0branchJ}{AiryF})
\beq
G_k(x)=\delta_{k0}\left(\pi n_i-\frac{1}{2x_0}\right)+
\left\{\!\!\!\!
\bea{rcl}
&\!\!\!\!\!\!c_k(x-x_0)^{\frac{1}{2}-\frac{3k}{2}}|a|^{\frac{1}{2}-
\frac{k}{2}}+\O\((x-x_0)^{1-\frac{3k}{2}}\), &\;\;\; a>0,\;x_0<0\\
(-1)^{k+1}&\!\!\!\!\!\!c_k(x_0-x)^{\frac{1}{2}-
\frac{3k}{2}}|a|^{\frac{1}{2}-\frac{k}{2}}+\O\((x_0-x)^{1-\frac{3k}{2}}\), &\;\;\; a<0,\;x_0>0
\eea
\right.\!\!\!,
\eeq
where the universal constants $c_k$ can be computed from the known asymptotic of Airy function
\beq
c_k=\left.\frac{{\rm Ai}'(z)}{{\rm Ai}(z)}\right|_{z^{-\frac{3k-1}{2}}},\;\;\;\;\;
{\rm Ai}(z)=\frac{e^{-\frac{2z^{3/2}}{3}}}{2\sqrt{\pi}z^{1/4}}\left[
\sum_{k=0}^n\frac{\left(\frac{1}{6}\right)_k\left(\frac{5}{6}\right)_k}{k!}\left(-\frac{3}{4z^{3/2}}\right)^k+\O\left(\frac{1}{z^{3(n+1)/2}}\right)
\right]
\eeq
in particular $c_0=-1,\;c_1=-\frac{1}{4},\;c_2=\frac{5}{32},\;c_3=-\frac{15}{64},\;c_4=\frac{1105}{2048},
\;c_5=-\frac{1695}{1024},\;c_6=\frac{414125}{65536},\;c_7=-\frac{59025}{2048}$.

These coefficients  behave asymptotically as $c_k\sim (-1)^k k!$ at $k\to\infty$.

We assume that $k\ll n,r$ and expand (for $x_0<0$)
\beq
\oint_{-y_0}^0(y+x_0)^{-n}y^\beta dy=
|x_0|^{\beta+1-n}(-1)^{n}\oint_{-y_0}^0 y^\beta e^{-n \log(1-y)}dy
\simeq|x_0|^{\beta+1-n}(-1)^{n}\oint_{-\infty}^0 y^\beta e^{ny}dy.
\eeq
For the last integral the path of integration starts at $-\infty-i0$, encircles the origin in the counterclockwise
direction, and returns to the point $-\infty+i0$. For the first integral the path is finite: it
starts at some point $-y_0-i0$
where $0<y_0<|x_0|$ and ends at $-y_0+i 0$. The dependence on the $y_0$
is exponentially suppressed. The last integral is nothing but the Hankel's contour integral
\beq
\oint_{-y_0}^0(y+x_0)^{-n}y^\beta dy=(-1)^n|x_0|^{\beta+1-n} n^{-\beta-1}\frac{2\pi i}{\Gamma(-\beta)}\left(1+\O\left(\frac{1}{n}\right)\right)
\eeq
similary
\beq
\oint_{0}^{y_0}(y+x_0)^{-n}(-y)^\beta dy=-|x_0|^{\beta+1-n} n^{-\beta-1}\frac{2\pi i}{\Gamma(-\beta)}\left(1+\O\left(\frac{1}{n}\right)\right)
\eeq
so that
\beq
\frac{G_k^{(n)}(0)}{n!}=
\left\{
\bea{rcl}
(-1)^{n}&\!\!\!\!\!\frac{c_k|a|^{\frac{1}{2}-\frac{k}{2}}n^{\frac{3k}{2}
-\frac{3}{2}}|x_0|^{\frac{1}{2}-\frac{3k}{2}-n}}{\Gamma\left(\frac{3k}{2}-\frac{1}{2}\right)}
\left(1+\O\left(\frac{1}{n}\right)\right),&\;\;\;\;\;a>0,\;x_0<0\\
(-1)^{k+1}&\!\!\!\!\!\frac{c_k|a|^{\frac{1}{2}-\frac{k}{2}}n^{\frac{3k}{2}
-\frac{3}{2}}|x_0|^{\frac{1}{2}-\frac{3k}{2}-n}}{\Gamma\left(\frac{3k}{2}-\frac{1}{2}\right)}
\left(1+\O\left(\frac{1}{n}\right)\right),&\;\;\;\;\;a<0,\;x_0>0
\eea
\right..
\eeq
As we can see from here,  only the term with $m=0$ in \eq{Qdecomp} contributes  at large $n$. The others are suppressed as $1/n$
and the final result is
\beq
Q_{k,r}=
\left\{
\bea{cl}
(-1)^{r}\frac{c_k |a|^{\frac{1}{2}-\frac{k}{2}}r^{\frac{3k}{2}-\frac{3}{2}}|x_0|^{\frac{3}{2}-\frac{3k}{2}-r}}{\Gamma\left(\frac{3k}{2}-\frac{1}{2}\right)}
\left(1+\O\left(r^{-1/2}\right)\right),&\;\;\;\;\;a>0,\;x_0<0\\
(-1)^{k}\frac{c_k |a|^{\frac{1}{2}-\frac{k}{2}}r^{\frac{3k}{2}-\frac{3}{2}}|x_0|^{\frac{3}{2}-\frac{3k}{2}-r}}{\Gamma\left(\frac{3k}{2}-\frac{1}{2}\right)}
\left(1+\O\left(r^{-1/2}\right)\right),&\;\;\;\;\;a<0,\;x_0>0
\eea
\right.,
\eeq
where we  introduced the notation
\beq
Q_r=\frac{1}{J^{r-1}}\sum_{k=0}^\infty Q_{k,r}\frac{1}{J^k}.
\eeq
Note that  $Q_{k,r}$ is  similar to $H_{g,n}$ of the matrix model.

\subsection{Non-local charges }

Now we will find the coefficients of the $1/x^n$ expansion for large $n$ and arbitrary $k$
\beq
G(x)=\sum_{k=0}^\infty\frac{1}{J^k}\sum_{n=1}^\infty\frac{d_{k,n}}{x^n},
\eeq
in other words
\beq
d_{k,n}=\frac{1}{2\pi i}\oint_{\cal C} x^{n-1}G_k(x)dx.
\eeq
In fact only the cut with minimal $n_i$ contributes for large $n$, or rather its branch point closest to $x=\infty$. The contributions of other branch points are exponentially suppressed. It is enough to consider only a small neighborhood
of the branch point with maximal $|x_{0,i}|$.
Near the branch point $p_k(x)$ will behave as
\beq\la{GnearBr2}
G_k(x)=\delta_{k0}\left(\pi n_i-\frac{1}{2x_0}\right)+
\left\{
\bea{rcl}
&\!\!\!\!\!\!c_k(x-x_0)^{\frac{1}{2}-\frac{3k}{2}}|a|^{\frac{1}{2}-\frac{k}{2}}+\O(x-x_0)^{1-\frac{3k}{2}}, &\;\;\;\;\; a>0,\;x_0>0\\
(-1)^{k+1}&\!\!\!\!\!\!c_k(x_0-x)^{\frac{1}{2}-\frac{3k}{2}}|a|^{\frac{1}{2}-\frac{k}{2}}+\O(x_0-x)^{1-\frac{3k}{2}}, &\;\;\;\;\; a<0,\;x_0<0
\eea
\right.
\eeq
similarly to the previous section we obtain
\beq
d_{k,n}=
\left\{
\bea{rcl}
&\!\!\!\!\!\!\frac{c_k\; a^{\frac{1}{2}-\frac{k}{2}}n^{\frac{3k}{2}-\frac{3}{2}}x_0^{n+\frac{1}{2}-\frac{3k}{2}}}{\Gamma\left(\frac{3k}{2}-\frac{1}{2}\right)}
\left(1+\O\left(n^{-1/2}\right)\right), &\;\;\;\;\; a>0,\;x_0>0\\
(-1)^{k+n-1}&\!\!\!\!\!\!\frac{c_k\; a^{\frac{1}{2}-\frac{k}{2}}n^{\frac{3k}{2}-\frac{3}{2}}x_0^{n+\frac{1}{2}-\frac{3k}{2}}}{\Gamma\left(\frac{3k}{2}-\frac{1}{2}\right)}
\left(1+\O\left(n^{-1/2}\right)\right), &\;\;\;\;\; a<0,\;x_0<0
\eea
\right..
\eeq
%

\section{ Conclusions}

We showed in this paper on the example of $\mathfrak{sl}(2)$ Heisenberg spin chain, how to find general solutions to quantum integrable problems in a specific thermodynamical limit proposed  by Sutherland, and how to find various finite size corrections to it. We also propose a double scaling analysis of the near edge distribution of Bethe roots giving some interesting results for the asymptotics of high conserved charges for the finite size corrections of any order.

Our methods  borrow some ideas from the matrix models: the size of the
chain $J$ is somewhat similar to the size of a random matrix $N$, and
the first two $1/J$  finite size corrections to the main limit which we calculated are of a similar mathematical nature as the $1/N$ corrections in matrix models. The asymptotics of high conserved charges found here in all orders of $1/J$ remind very much the double scaling approach in  matrix models
\cite{BrezinKazakov,Douglas:1989ve,Gross:1989vs}. To calculate them we used for that
the Airy asymptotics on the edge of the Bethe root distribution, similar
 to the generic edge behavior in the matrix models.  However, it is obvious
  that by finetuning the parameters of the chain and of its large $J$
   solutions we can reach various multicritical points with a different
    from Airy classes of universality, probably also similar to those
     of the multicritical matrix models \cite{Kazakov:1989bc} or a
     two cut model \cite{Periwal:1990gf,Douglas:1990xv}. Probably,
     some modern methods of analysis of the doubly scaled matrix models,
     like the Riemann-Hilbert method for the asymptotics of orthogonal polynomials \cite{Bleher:2002ys}
     should work as well for the quantum integrable chains in this thermodynamical limit. Indeed, it is known that the
polynomials $Q_S(u)=\prod_{k=1}^S(u-u_k)$ are orthogonal for different $S$
with a specific $S$-independent measure \cite{VarchenkoMukhin}.

Our methods might be useful for finding the asymptotics of numbers of various combinatorial objects on regular lattices related to the integrable models, such  the tilting patterns, spanning trees, dimers etc., as its analogue was useful for finding the asymptotics of various large planar graphs via the matrix models.

The methods presented here can be easily carried over to the $\mathfrak{su}(2)$ quantum
chain as well, though some peculiarities of this model, like
 complex distributions of roots and the presence of "string"
  condensates  with equally distributed roots \cite{Beisert:2003xu},
   should be taken into account. Only slight modifications of our results will allow
 to find the $1/J$ corrections in the nonlocal integrable deformations
of the $\mathfrak{su}(2)$ spin chain described in \cite{Beisert:2003tq,Serban:2004jf}.
As for  more complicated models solved by nested Bethe ansatz and with the thermodynamical limit described by non-hyperelliptic algebraic curves, the $1/J$ and $1/J^2$ correactions are left to be established.

In the context of integrability in the 4D Yang-Mills theory with ${\cal N}=4$
supersymmetries (SYM), where the hamiltonian of the $\mathfrak{psu}(2,2|4)$
integrable spin chains describes the matrix of anomalous dimensions, our methods could be especially useful.
The appropriate thermodynamical limit in terms of  non-hyperelliptic
curve, corresponding to very long operators in SYM  was constructed
in particular sectors in  \cite{KMMZ,Kazakov:2004nh,Beisert:2004ag,Schafer-Nameki:2004ik}
 and for the full $\mathfrak{psu}(2,2|4)$ SYM chain in \cite{Beisert:2005di}.
Finding the finite size corrections to these solutions might
 be extremely useful for getting some clues for the quantization
  of the string sigma model on the other side of  AdS/CFT correspondence.
The finite size correction for the simplest $\mathfrak{su}(3)$   solution of
SYM theory found in \cite{Engquist:2003rn} are not available yet, let alone
 more complicated solutions and sectors. The finite size corrections
  to this solution, if found, could be compared  to the string quantum
   correction computed in \cite{Frolov:2004bh}. The corrections found
    here for the most general  multi-cut solution in $\mathfrak{sl}(2)$ sector
    could shed some light on the structure of the first quantum corrections to string
     solitons describing spinning strings, where for the moment only the direct,
     very cumbersome quasi-classical methods of evaluation of the functional integral near the simplest classical solutions
      are known \cite{FrolovTseytlin}.

As for the double scaling limit used here for the refined analysis near the edge of a distribution and giving the asymptotics of high conserved charges to any order of $1/J$, it is very universal and is not based on any particular form of solutions. Near the edge we can forget about the rest of the sheets and cuts in any algebraic curve, and the asymptotics will be governed in all cases by the Airy functions. In fact, we think that this near edge behavior will persist also in the near-classical regime of the string theory on $AdS_5\times S^5$, and even for some of its recently considered deformations. We expect then that the asymptotics of high conserved charges at all orders of $1/J$ will be also dominated by the same Airy type solution as found here. It is already a deeply quantum regime (reminding that of the non-critical string theories in $\le 2$ dimensions) and it could be extremely useful as a step to the full quantization and solution  of the Metsaev-Tseytlin supersgtring and hence of the ${\cal N}=4$ SYM theory.

\subsection*{Acknowledgements}

We would like to thank N.~Beisert, I.~Kostov,  F.~Smirnov and K.~Zarembo   for discussions.
The work of V.K. was partially supported
by European Union under the RTN contracts MRTN-CT-2004-512194 and by INTAS-03-51-5460 grant.
The work of
N.G. was partially supported by French Government PhD fellowship and by
RSGSS-1124.2003.2.
We also thank Institut Universitaire de France for a partial support.

\section*{Appendix A, BAE $1/J$ expansion}

In this Appendix we give the formulas for the expansion
of the rhs of BAE  (\ref{BAE}) in powers of $1/J$.
The density $\vrho(x)$ is defined in \eq{defvro}.
We assume $k, S-k\sim J$ (i.e. far from the ends of the
cut)
\beqa
&&{\sum_j}'i\log\left(\frac{u_j-u_k+i}{u_j-u_k-i}\right)
 \simeq -2{\sum_j}'\frac{1}{u_j-u_k}+
\frac{\pi\vrho'[\coth(\pi\vrho)]_0}{J}+
\O\left(\frac{1}{J^{2}}\right)\\
\nn&&{\sum_j}'i\log\left(\frac{u_j-u_k+i}{u_j-u_k-i}\right)
 \simeq -2{\sum_j}'\frac{1}{u_j-u_k}+
\frac{2}{3}{\sum_j}'\frac{1}{(u_j-u_k)^3}+
\frac{\pi\vrho'[\coth(\pi\vrho)]_2}{J}+
\O\left(\frac{1}{J^{3}}\right)\\
\nn&&{\sum_j}'i\log\left(\frac{u_j-u_k+i}{u_j-u_k-i}\right)
 \simeq -2{\sum_j}'\frac{1}{u_j-u_k}+
\frac{2}{3}{\sum_j}'\frac{1}{(u_j-u_k)^3}
-\frac{2}{5}{\sum_j}'\frac{1}{(u_j-u_k)^5}+
\frac{\pi\vrho'[\coth(\pi\vrho)]_4}{J}\nn\\
\nn&&-\frac{1}{12 J^3}\left((\pi\vrho')^3
\left[\frac{\coth(\pi\vrho)}{\sinh^2(\pi\vrho)}
\right]_0-2\pi^2\vrho'\vrho''
\left[\frac{1}{\sinh(\pi\vrho)}\right]_1+
\pi\vrho'''[\coth(\pi\vrho)]_2\right)+
\O\left(\frac{1}{J^{4}}\right) \\
&&{\sum_j}'i\log\left(\frac{u_j-u_k+i}{u_j-u_k-i}\right)
 \simeq -2{\sum_j}'\frac{1}{u_j-u_k}+
\frac{2}{3}{\sum_j}'\frac{1}{(u_j-u_k)^3}-
\frac{2}{5}{\sum_j}'\frac{1}{(u_j-u_k)^5}+
\frac{2}{7}{\sum_j}'\frac{1}{(u_j-u_k)^7}\nn\\
&&+\nn
\frac{\pi\vrho'[\coth(\pi\vrho)]_6}{J} -\frac{1}{12
J^3}\left((\pi\vrho')^3
\left[\frac{\coth(\pi\vrho)}{\sinh^2(\pi\vrho)}\right]_2
-2\pi^2\vrho'\vrho''
\left[\frac{1}{\sinh(\pi\vrho)}\right]_3
+\pi\vrho'''[\coth(\pi\vrho)]_4\right)
+\O\left(\frac{1}{J^{5}}\right),
\eeqa
where we introduced the notation  $[f(\vrho)]_n\equiv f(\vrho)-\sum_{i=0}^{n-1} f^{(i)}(0)\frac{\vrho^i}{i!}$
for the functions regular at zero. For singular functions the Taylor series should be substituted by the Laurent series, so that
$[f(\vrho)]_n$ is zero for $\vrho=0$ and has first $n-1$ zero derivatives at this point.
For example
$[\coth(\pi\vrho)]_2\equiv
\coth(\pi\vrho)-\frac{1}{\pi\vrho}-\frac{\pi\vrho}{3}$.

\section*{Appendix B, Example: 1-cut}

In this Appendix we express corrections to the
energy in terms of infinite sums for the simplest
case of one-cut solution.
For this solution the hyperelliptic curve is a sphere.
It is two complex planes connected by a single cut. The
density of the Bethe roots is given by a simple formula \cite{Kazakov:2004nh}
\beq
\rho(x)=\frac{\sqrt{8\pi m x-(2\pi n x-1)^2}}{2\pi x}.
\eeq
We can easily find explicit expressions for $a_i$ and $b_i$ of \eq{p2PL}. With the
notation $M=\sqrt{m(m+n)}$ $a_i$ and
$b_i$ become
\beq \la{a1b1}a_1=-\frac{8M n^4
\pi^3}{(\sqrt{4M^2+n^2}-2M)^2},\;\;\;\;\;b_1
=\frac{4\pi^4n^6}{3(\sqrt{4M^2+n^2}-2M)^4}
\left(12M\sqrt{4M^2+n^2}+3n^2-4n^2\pi^2M^2-24M^2\right)
\eeq
and
\beq \la{a2b2}a_2=\frac{8M n^4
\pi^3}{(\sqrt{4M^2+n^2}+2M)^2},\;\;\;\;\;b_2=
-\frac{4\pi^4n^6}{3(\sqrt{4M^2+n^2}+2M)^4}
\left(12M\sqrt{4M^2+n^2}-3n^2+4n^2\pi^2M^2+24M^2\right) .\eeq
It may
be more convenient for comparison with string theory results \cite{FrolovTseytlin} to express $A$ defined by
\eq{p0branchP} as an
infinite sum. We have to evaluate the integral in \eq{p1res} and find $A$ from
the behavior near a branch point. We compute the integral by poles.
To that end we use that the solutions to the equation $\sin(p_0(x_l^{\pm}))=0$ are
\beq
x_l^\pm=\frac{1}{2\pi}
\frac{1}{\sqrt{4M^2+n^2}\mp\sqrt{4M^2+l^2}},\;\;\;\;\;l\geq
0 .\eeq
The  points $x^\pm_{l=0}$ are the branch points. They are inside the
contour of integration and thus do not contribute.

Using that $f(x_l^\pm)/x_l^\pm=\pm \frac{l}{n}$ and
\beqa
\frac{1}{x^+_l-x_{0,1}}-\frac{1}{x^-_l-x_{0,1}}&=&
-\frac{\sqrt{l^2+4M^2}}{l^2}\frac{1}{\pi x_{0,1}^2}\\
\nn\frac{1}{x^+_l-x_{0,2}}-\frac{1}{x^-_l-x_{0,2}}&=&
-\frac{\sqrt{l^2+4M^2}}{l^2}\frac{1}{\pi x_{0,2}^2}.
\eeqa
We can evaluate the integral \eq{p1res} for $x\rightarrow x_{0}$ (we also take $x$ inside the contour
to drop irrelevant symmetric part of $p_1$)
\beq
\frac{1}{2\pi i}\oint_{{\cal C}}\frac{f(y)p'(y)\cot
p(y)}{y(y-x)}dy\rightarrow
-\frac{1}{i\pi n x_0^2}\sum_{l=1}^\infty\frac{\sqrt{l^2+4M^2}}{l}
\eeq
we can conclude that
\beqa
\la{A1cut}A_2&=&-\frac{1}{2 x_2^2\sqrt{a_2}}\sum_{l=1}^\infty\frac{\sqrt{l^2+4M^2}}{l}\\
\nn A_1&=&-\frac{1}{2
x_1^2\sqrt{-a_1}}\sum_{l=1}^\infty\frac{\sqrt{l^2+4M^2}}{l}\;, \eeqa
where the sum should be understood in the zeta-function
regularization (a natural explanation why this regularization
gives the right result is given in \cite{Beisert:2005mq}. A
more regular way to express the integral as a sum is to
expand $\cot$ into the sum before integration)
\beq
\sum_{l=1}^\infty\frac{\sqrt{l^2+4M^2}}{l}\equiv
\sum_{l=1}^\infty\left(\frac{\sqrt{l^2+4M^2}}{l}-1\right)-\frac{1}{2}.
\eeq
We can easily reproduce the result of \cite{Beisert:2005mq} for $E_1$ in terms of a sum from \eq{E1fnl}
\beq
E_1=-p'_1(0)=4\pi^2\sum_{l=1}^\infty l\sqrt{l^2+4 M^2}
\eeq
with $\zeta$-function regularization assumed.

We can also express our result for the next correction to the
energy $E_2$ given by \eq{ETWO} as a double sum. We will need the following
quantity
\beq
p_1(x_k^\pm)=\frac{\pm1}{2\pi(x_k^\pm)^2k}\left[\sum_{l=1}^\infty\left(\frac{l\sqrt{l^2+4 M^2}-k\sqrt{k^2+4 M^2}}{l^2-k^2}-1\right)
+\frac{\sqrt{k^2+4M^2}}{2k}-\frac{1}{2}\right].
\eeq
Evaluating the integrals in \eq{ETWO} we  express $E_2$ as a double sum
\beq
E_2=-({\cal I}_1+{\cal I}_2+{\cal I}_3+{\cal I}_4),
\eeq
where
\beqa
\nn{\cal I}_1&\equiv&\frac{1}{4\pi i f(0)}\oint\frac{f(z)}{z^2}\d_z(p_1\cot p_0)=-2p_1'(0)+
\sum_{k=1}^\infty\left[2\pi\sum_\pm \left(\sqrt{4M^2+n^2}\pm 2\frac{k^2+2M^2}{\sqrt{k^2+4M^2}}\right)p_1(x_k^\pm)-4p_1'(0)\right]\\
{\cal I}_2&\equiv&\frac{1}{4\pi i f(0)}\oint\frac{f(z)}{4z^5}=4\pi^4 M^2(n^2+5M^2)\\
\nn{\cal I}_3&\equiv& I_2'(0)=\frac{1}{16}\left(\frac{1}{x_{0,1}^2}+\frac{1}{x_{0,2}^2}\right)
+\frac{1}{x_{0,1}}\left(\frac{7a_1}{96}-\frac{b_1}{8a_1}-A_1^2\right)
+\frac{1}{x_{0,2}}\left(\frac{7a_2}{96}-\frac{b_2}{8a_2}+A_2^2\right)\\
\nn{\cal I}_4&\equiv&-\frac{G_0''(0)f'(0)}{24 f(0)}-\frac{G_0^{(3)}(0)}{24}=\frac{4}{3}M^2(2n^2+11M^2)\pi^4.
\eeqa
Note that in our new notations
$1/x_{0,i}=4\pi M\pm2\pi\sqrt{4M^2+n^2}$.
Expressions for $a_i,\;b_i$ and $A_i$ are given in \eqs{a1b1}{a2b2} and \eq{A1cut}.


\end{document}